\documentclass[conference]{IEEEtran}
\IEEEoverridecommandlockouts
\usepackage{cite}
\usepackage{algpseudocode} 
\usepackage{colortbl}
\definecolor{lightgray}{gray}{0.7}
\usepackage{amsmath,amssymb,amsfonts}
\usepackage{bbm}
\usepackage{mathrsfs}
\usepackage{dutchcal}
\usepackage{graphicx}
\usepackage{textcomp}
\usepackage{xcolor}
\usepackage{threeparttable}
\allowdisplaybreaks
\def\BibTeX{{\rm B\kern-.05em{\sc i\kern-.025em b}\kern-.08em
    T\kern-.1667em\lower.7ex\hbox{E}\kern-.125emX}}
\begin{document}

\title{Optimal Mobility Aware Wireless Edge Cloud Support for the Metaverse }
\author{\IEEEauthorblockN{Zhaohui Huang and Vasilis Friderikos}
\IEEEauthorblockA{Center of Telecommunication Research, 
King's College London,
London, U.K. \\
E-mail: \{zhaohui.huang, vasilis.friderikos\} @kcl.ac.uk}
}

\maketitle

\begin{abstract}
Mobile augmented reality (MAR) applications extended in the metaverse could provide mixed and immersive experiences by amalgamating the virtual and physical world. However, the joint consideration between MAR and metaverse seeks the reliable and high quality support for foreground interactions and background contents from these applications, which intensifies their consumption in energy, caching and computing resources. To tackle these challenges, a more flexible request assignment and resource allocation with more efficient processing are proposed in this paper through anchoring decomposed metaverse AR services at different edge nodes and proactively caching background metaverse region models embedded with target Augmented Reality Objects (AROs).       
% However, such applications are intense in terms of energy consumption, computational and caching resources to support foreground interactions and background contents. In this paper, the metaverse MAR service is decomposed and anchored at different edge nodes to enable efficient processing of background metaverse region models embedded with target Augmented Reality Objects (AROs).
 Advanced terminals are also considered to further reduce service delay at an acceptable cost of energy consumption.
 We then propose and solve a joint optimization problem that explicitly considers the balance between service delay and energy consumption under the constraint of user perception quality in a mobility event. 
 % To achieve that, a joint optimization problem is proposed, which explicitly considers the user physical mobility, service decomposition, and the balance between service delay and energy consumption under the constraint of user perception quality. 
 By also explicitly taking into account capabilities of user terminals, the proposed optimized scheme is compared to its terminal oblivious version in this paper. According to a wide set of numerical investigations, the proposed scheme owns advantages in service latency and energy efficiency over other nominal baseline schemes which neglect capacities of terminals, user physical mobility, service decomposition and the inherent multi modality of the metaverse MAR service.  
\end{abstract}

\begin{IEEEkeywords}
Metaverse, 5G, Augmented Reality, Mobility, Structural Similarity (SSIM), Energy Consumption 
\end{IEEEkeywords}
\setlength{\parskip}{0em}

\section{Introduction}
\IEEEPARstart{M}{obile} Augmented reality (MAR) could be extended and enhanced in wireless edge supported metaverse by today's available technologies like digital twin and head-mounted display rendering \cite{xu2021wireless}\cite{dong2019deep}. Compared to ongoing MAR applications, users could have mixed experience seamlessly between metaverse and physical world through various metaverse MAR applications like massively multiplayer online video games and virtual concerts \cite{xu2021wireless}. Users equipped with MAR devices can upload and analyze their environment through AR customization to achieve appropriate AR objects (AROs) and access the metaverse in mobile edge networks \cite{xu2022full}. Rendering 3-dimensional (3D) AROs with the background virtual environment and updating in metaverse consume significant amounts of energy and could be highly demanding in terms of required caching and computing resources \cite{xu2022full}\cite{li2020rendering}. Hence, such applications are delay and energy sensitive and face challenges in ensuring user quality of experience and providing reliable and timely interactions with the metaverse \cite{xu2021wireless}\cite{xu2022full}.  

\begin{figure}[htb]
\centering
\includegraphics[width=0.95\linewidth]{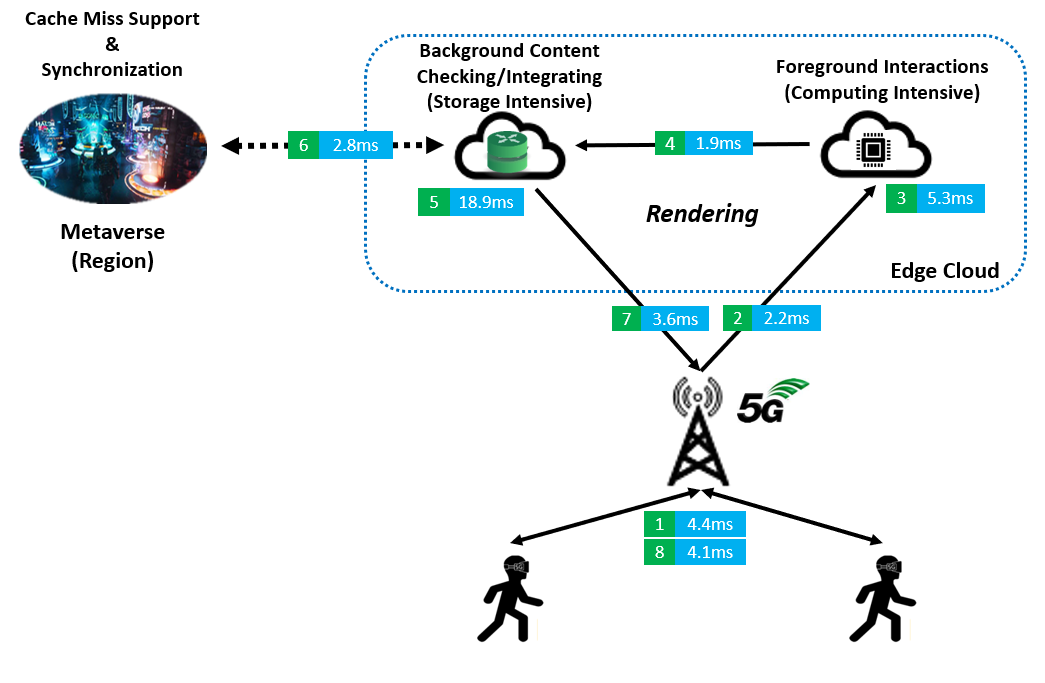}
\caption{The general work flow of metaverse AR applications with delay of each stage (6 EC, 30 requests, weight $\mu$ is 1, EC Capacity is 14 and total mobility probability is 1)}
\label{fig:workflow}
\end{figure}

Generally speaking, a metaverse scene will in essence consisted by a background view as well as many objects in foreground interactions. The background view at a defined amalgamated virtual and physical location cane be deemed as static or slowly changing \cite{xu2022full}\cite{guo2020adaptive}. A typical background scene can be the 3D model of the metaverse, a presentation of a related background virtual environment based on a certain user viewport \cite{guo2020adaptive}\cite{kato2021split}. Its size can reach tens of MB and the corresponding complexity of rendering related functionalities measured by computation load is also large (e.g., 10 CPU cycles/bit) \cite{guo2020adaptive}\cite{yang2018communication}. On the other hand, objects (such as for example avatars) in foreground interactions that are embedded in the metaverse scene change much more frequently however they are significantly less complex than the background scene (e.g., 4 CPU cycles/bit) \cite{li2020rendering}\cite{guo2020adaptive}. But, even though those objects are less complex than the background scene, due to their frequent changes they also require rendering them in a timely manner so that to avoid a considerable quality of experience degradation. Thus, in this paper, rendering for both foreground and background are deployed at the edge clouds (ECs) rather than only at the terminals to make full use of the caching and computing resources. Noticing that uploaded information are focused in foreground interactions while background content checking consumes not only computing resources but also lots of local cache to match and integrate AROs and related models of the metaverse. Hence, similar to our previous work in \cite{huang2021proactive}, the metaverse AR application could also be decomposed into computational and storage intensive functions which serve as a chain for improved assignment and resource allocation. 

The general work flow of a metaverse MAR application supported by ECs is shown by Fig. \eqref{fig:workflow}. A metaverse region is a fraction of the complete metaverse and is assumed to be stored on a server geographically close to its corresponding service region in the mobile network (not necessarily running within an EC). The metaverse AR service could be triggered by certain behavior with foreground interactions \cite{xu2022full}\cite{guo2020adaptive}\cite{hertzmann2000painterly}. Then, content related to background contents like for example pre-cached 3D models and AROs are firstly searched in the EC cache to check if they are what the user requires. If the target AROs or model information cannot be found in the cache, then this case is labelled as a "cache miss" and the request is redirected to the original metaverse region stored in a cloud deeper in the network. Finally, according to user's physical mobility and virtual orientation extracted from foreground interactions, the matched AROs and model are integrated into a final frame and transmitted back to the user \cite{xu2022full}\cite{guo2020adaptive}. At the same time, updated information is also sent to the metaverse region for synchronization. Thus, the user could be aware of changes caused by other participants if they share the same metaverse region during the service. The overall quality of the metaverse AR application depends on the communication delays and the capabilities of the above mentioned network entities that participate in the service creation. Hereafter, we assume a nominal frame rate as 15 frames/second and the rendering happens at every other frame (\~ 133.2ms interval) \cite{cozzolino2022nimbus}\cite{niu2018learning}\cite{naman2013inter}. Thus, the service delay of the aforementioned work flow within the interval could be regarded as acceptable. 
\begin{figure}
    \centering
    \begin{minipage}[]{0.9\linewidth}
     \centering
    \includegraphics[width=1\linewidth]{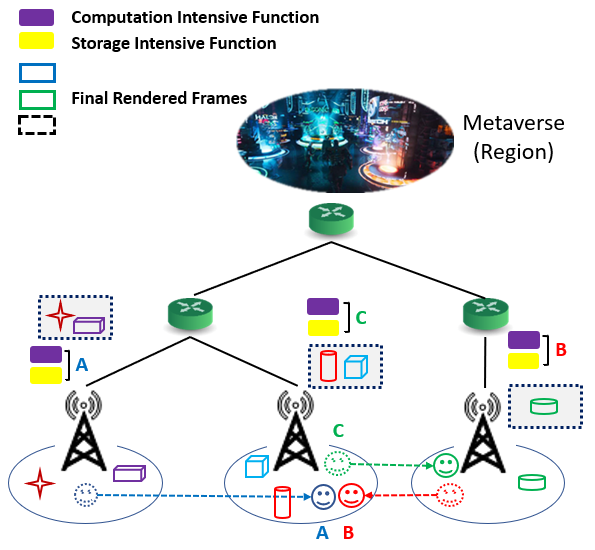}
    \end{minipage}
    \begin{minipage}[]{0.9\linewidth}
    \centering
    \includegraphics[width=1\linewidth]{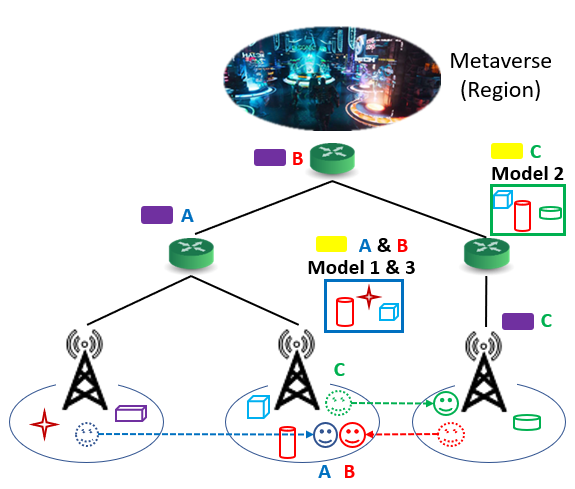}
    \end{minipage}
    \caption{Illustrative toy examples of a metaverse MAR application where in case (a) user mobility is not considered and in case (b) the physical mobility of the end user and service decomposition are considered with different renderings on pro-active resource allocation of the metaverse MAR functions.}
    \label{fig: toyMetaAR}
\end{figure}

\begin{figure}[htb]
\centering
\includegraphics[width=0.98\linewidth]{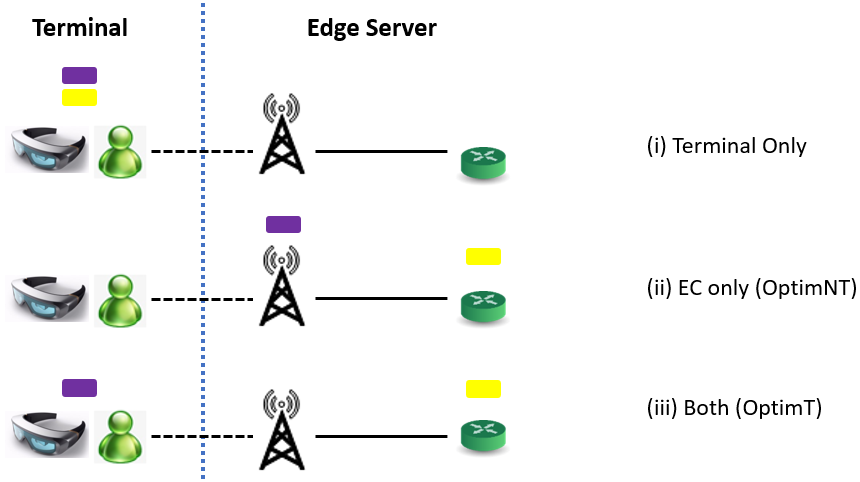}
\caption{Illustrative toy examples where in case (i) only terminals are activated, in case (ii) only ECs are activated and in case (iii) both are activated}
\label{fig:TerminalEC}
\end{figure}
Fig. \eqref{fig: toyMetaAR} further reveals the difference between cases that consider the user mobility with service decomposition or not related to rendering requirements in metaverse AR applications. Clearly, when neglecting user mobility and service decomposition as shown by case (a), models, target AROs and metaverse AR applications are all cached close to the user's initial location. It might leave a heavy burden for the adjacent server when there are multiple users at the same cell and causes a "hot spot" area \cite{huang2021proactive}. However, when user mobility and service decomposition is enabled, as shown in case (b), then service delivery becomes more flexible and efficient in terms of assigning requests and allocating network resources. In case (a), although the user A only needs wireless communication in the initial location, it takes 2 hops after moving to the middle cell. However, when taking mobility and local resources into consideration, the same user A in case (b) could encounter a better delay in the after mobility event by allowing 2 more hops in the before mobility event. Hence, in a high mobility scenario, it might not always be ideal to allocate requests and services as close as possible to the user's initial location. As shown in the Figure for users A and B in case (b), the AR contents in the model might be similar in terms of the viewport of different users. Hence, participating users should be aware of each other's updates and could share rendering functions to reduce the consumed resources. In this paper, we apply Structural Similarity (SSIM) proposed by \cite{wang2004image} for user perception experience. It is a widely accepted method that measures user perception quality of an image by comparing to its original version \cite{wang2004image}. Caching more models and AROs also causes more processing and transmission delay with energy consumption \cite{xu2021wireless}\cite{xu2022full}. Hence, the the joint optimization has to accept some potential loss due to constraints of computing and storage resources. Fig. \eqref{fig:TerminalEC} further reveals the difference among cases whether allocating MAR service on terminals or ECs. Neglecting the EC support and service decomposition, the whole MAR application is left to the terminal and could be a heavy burden in case (i). According to \cite{xu2022full}\cite{chen2018understanding}, MAR application on terminals takes up around 47\% power consumption and could affect the running of other functionalities when this service becomes more complex like in metaverse. In case (ii) that researched in our previous work \cite{huang2022mobility}, the processing time is significantly cut down through enabling ECs with the cost of extra transmission delay. Although the terminals are still limited in computing resources and more sensitive in energy consumption, neglecting their capacity at all seems also a waste, especially when recent technical improvements approve their computing and caching potential. Noticing that the foreground interactions are much less complex that background scenes and are more suitable to terminals, we further bring in terminals into the scenario as shown by case (iii) and executing only computational intensive functions. In this paper, the optimization under the integration of terminals and ECs becomes the scheme OptimT and is compared to the previous case (ii) neglecting the terminals (OptimNT). To maintain a fair comparison and focus more on energy consumption and service latency, the overall energy in Joule is measured instead of power in \cite{huang2022mobility} and the user perception quality is accepted as a given boundary.

In this paper, by considering explicitly the terminals, user mobility, service decomposition and models of metaverse regions with embedded AROs, a joint optimization framework (OptimT) is constructed for the metaverse MAR application in the edge supported network. The proposed optimization framework seeks a balance between energy consumption and service delay under a given level of user perception quality. To reveal the influence of capacity and cost of terminals on the metaverse MAR application, this OptimT scheme is further compared with another optimized framework (OptimNT) proposed in our previous work \cite{huang2022mobility} that only focuses on ECs and neglects terminals.

\section{Related Work}
Hereafter, a series of closely related works in the area edge/cloud support of metaverse type of applications over 5G and beyond wireless networks are discussed and compared with the approach proposed in this paper.

Noticing that MAR terminals have witnessed a series of technical improvements while are still quite time and computational consuming, their utilization are focused and accepted as part of the network optimization. In \cite{song2019energy}, authors try to minimize the energy consumption of multicore smart devices, which are usually applied for AR applications. Through tracking the response process of an AR application, they manage to measure the terminal's energy consumption by Amdahl's law. Although the law and a similar formula for the terminal energy are also applied in this paper, we consider a larger scenario including edge servers and a more complex balance between energy and latency. \cite{chen2018marvel} shares a similar target like us, which is achieving the balance between latency and energy consumption under an level of acceptable image quality. However, \cite{chen2018marvel} brings in local sensors at MAR devices for recognizing and tracking AROs so that their scheme could realize selective local visual tracking (optical flow) and selective image offloading. Object recognition stage is more focused in \cite{chen2018marvel} with four AR applications requiring different types of AROs. Without support from ECs, they utilize a further cloud server as a heavy database for 3D AROs and leave most tasks at terminals and cloud offloading only triggers when confronted with a calibration. While in this work, we accept an EC supported network and compare the difference between schemes utilizing MAR terminals or not. In \cite{seo2021novel}, the energy efficiency is optimized under required service latency for MAR in an EC supported network. Similarly, authors consider the proactively caching and propose a tradeoff between energy and latency in terms of cache size. However, their mobile cache and power management scheme still focuses the energy consumption on terminals. Clearly, above mentioned works do not explicitly consider user mobility, perception quality, service decomposition and metaverse application features like ours. 

The problem of efficient resource allocation for supporting metaverse type of applications is starting to attract significant amount of attention and various aspects have been considered. In \cite{han2021dynamic}, the emphasis is placed  in the  synchronization of Internet of Things (IoT) services, in which they employ IoT devices to collect real world data for virtual service providers. Through calculating maximum awards, users could select the ideal virtual service provider. Researchers then proposes the game framework considers such reward allocation scheme and general metaverse sensing model \cite{han2021dynamic}. In  \cite{jiang2021reliable}, the authors also adopt a game theoretic framework by considering tasks offloading between mobile devices based on Coded Distributed Computing (CDC) in a proposed vehicular metaverse environment. Another framework proposed by \cite{chu2022metaslicing} manages and allocates different types of metaverse applications so that common resources among them could be shared through a semi-Markov decision process and an optimal admission control scheme. The work in \cite{ng2021unified} applies a set of proposed resource optimization schemes in a virtual education metaverse. Mores specifically, a stochastic optimal resource allocation
scheme is developed with the aim to reduce the overall cost of incurred by a service provider. Similar to the service decomposition in this paper, they only upload and cache some parts of data or services to achieve reduced levels of delay and offer better privacy \cite{ng2021unified}. The work in \cite{dong2019deep} is closely relevant since in that paper not only latency but energy consumption is also considered as is the case of our proposed model that uses a multi-objective optimization approach. For ultra-reliable and low-latency communication services, researchers bring in digital twins and deploy a mobility management entity for each access point to determine probabilities of resource allocation and data offloading \cite{dong2019deep}. Then, by applying a deep learning neuro network the proposed scheme tries to identify a suitable user association and an optimized resource allocation scheme for this association. While, in this paper, the core idea is to decompose he service and allow a flexible allocation across edge clouds by taking also into account user mobility.  The work in\cite{xu2021wireless} considers virtual reality applications in the metaverse and regards the service delivery as a series of events in the market, in which users are buyers and service provides are sellers. Hence, they apply Double Dutch Auction (DDA) to achieve common price through asynchronous and iterative biding stages \cite{xu2021wireless}. They emphasize the quality of user perception experience by Structural SIMilarity (SSIM) and Video Multi-Method Assessment Fusion (VMAF). In our proposed framework, we also utilize the SSIM metric to determine the frame quality after integrating background scene and AR contents \cite{xu2021wireless}. The work in \cite{xu2021wireless} further brings in a deep reinforcement learning-based auctioneer to reduce the information exchanging cost. While in this paper, a multi-objective optimization approach is taken where we aim to balance across different objectives using scalarization and considering user mobility in an explicit manner.

\section{System Model}
\subsection{Multi Rendering in Metaverse AR}
In a given wireless network topology, the set $\mathbb{M}=\{1,2,...,M\}$ denotes the available locations, including available edge clouds and terminals. Assuming that each user makes a single request, the corresponding MAR service requests defined by $r\in \mathbb{R}$ in the metaverse region generated by mobile users could be equipped with MAR devices. Request $r$ emerge from network location $f(r)$. It represents the initial access router where this user is firstly connected to. For each request, the user terminal could also be viewed as a valid location to cache or to process information locally. Thus, we define with $j_r \in \mathbb{M}$ as the terminal sending the request $r$. The terminals are brought into consideration in the following formulation. Through defining the location with constraint as $j\in \mathbb{M}, j\neq j_r$, this could only enable the ECs. Clearly, for the OptimNT scheme, this works for all locations. While in the following formulation for the OptimT scheme, we force that the storage intensive functions are executed only on the ECs while the computing intensive ones could also hosted at the end terminals. During the mobility event, a user could move to different potential destinations $k\in \mathbb{K}$ (i.e., changing of the anchoring point). Hereafter, and without loss of generality, we only accept adjacent access routers as available destinations in the mobility event. A series of metaverse regions are set on ECs to interact with users. The corresponding metaverse region serving the user can be found through functions $A(f(r)), A(k)$. As explained earlier, each metaverse region is pre-deployed on a server close to the mobile network and its distance to an EC is also predefined. In this paper, as already alluded, a set of AROs is assumed to be embedded across the different background metaverse region models and is defined as $\mathbb{N}=\{1,2,...,N\}$. A set $\mathbb{S}_r=\{1,2,...,S\}$ is defined for multiple rendering of the available metaverse region model to each user. Thus, we denote the decision variable $p_{sj}$ for pre-caching a metaverse region model $s\in \mathbb{S}_r$ at the EC $j$ ($j\in \mathbb{M}, j \neq j_r$). The subset $\mathbb{L}_{rs}$ represents the target AROs required by the user $r$ in the related model $s\in \mathbb{S}_r$ and the size of each target ARO $l\in \mathbb{L}_{rs}$ is denoted as $O_l$. Lastly, the decision variable $h^s_{rl}$ is brought in for proactively caching an ARO required by a request $r$. Based on the above, the decision variables $p_{sj}$ and $h^s_{rl}$ can be defined as follows,
\begin{equation}
p^{j}=\left\{
\begin{aligned}
1,&\;\text{if rendering the related model}\;s\;\text{at  node}\;j, \\
0,&\;\text{otherwise}.
\end{aligned}
\right.
\end{equation}
\begin{equation}
h^s_{rl}=\left\{
\begin{aligned}
1,&\;\text{if ARO}\;l\;\text{required by request}\; r\;\text{embedded}\qquad\qquad\qquad\quad\\&\;\text{in the model}\;s\;\text{is cached}, \\
0,&\;\text{otherwise}.
\end{aligned}
\right.
\end{equation} 
In addition to the above, the following constraints should also be satisfied,
\begin{equation}
\begin{aligned}
\sum_{r \in \mathbf{R}}h^s_{rl} \leqslant 1,\; \forall j\in \mathbf{M}, j\neq j_r,\;\forall s\in \mathbf{S_r},\; \forall l\in \mathbf{L_{rs}} \\
\end{aligned}
\label{cons_h1}
\end{equation}
\begin{equation}
\begin{aligned}
\sum_{s\in \mathbf{S_r}}\sum_{l \in \mathbf{L_{rs}}}h^s_{rl} \geqslant 1,\; \forall r\in \mathbf{R} \\
\end{aligned}
\label{cons_h2}
\end{equation} 
\begin{equation}
\begin{aligned}
\sum_{j \in \mathbf{M}, j\neq j_r }p_{sj} \geqslant h^s_{rl},\; \forall r\in \mathbf{R},\;\forall s\in \mathbf{S_r},\; \forall l\in  \mathbf{L_{rs}} \\
\end{aligned}
\label{cons_h3}
\end{equation}
\begin{equation}
\begin{aligned}
h^s_{rl} \leqslant h^s_{rl} \sum_{j \in \mathbf{M}, j\neq J_r}p_{sj},\; \forall r\in \mathbf{R},\;\forall s\in \mathbf{S_r},\; \forall l\in  \mathbf{L_{rs}} \\
\end{aligned}
\label{cons_h4}
\end{equation}
Constraints in \eqref{cons_h1} forces each ARO to be pre-cached at most once in a related model. In constraints \eqref{cons_h2}, a valid request consists of at least one model and an embedded ARO. Constraints in \eqref{cons_h3} guarantee that the allocation of an ARO happens in conjunction to the decision of proactive caching whilst constraints in \eqref{cons_h4} further certify that the rejection of the model's proactive caching causes any ARO planned to be embedded in this model should not be pre-cached as well. Thus, \eqref{cons_h3} only accepts ARO in an pre-cached model and \eqref{cons_h4} rejects all related ones when failing to cache a model, which together could ensure the model and corresponding AROs cannot be handled separately during the formulation.

\subsection{Wireless Channel Model}
With $B_j$ we express the bandwidth of the resource block and $\gamma_{rj}$ denotes the Signal to Interference plus Noise Ratio (SINR) of the user $r$ at the node $j$. With $P^{tran}_{rj}$ we denote the transmit power of the user $r$ at the node $j$, and $P_i$ is transmission power at the base station. Furthermore,  $H_{rj}$ is the channel gain, $N_j$ is the noise power and $a$ is the path loss exponent whilst $d_{rj}$ is the  distance between the user and the base station. Finally, a  Rayleigh fading channel is used to capture the channel between the base stations and the users\cite{gemici2021modeling}. More specifically, the channel gain $H_{rj}$ can be written as follows \cite{cho2010mimo}, 
\begin{equation}
\begin{aligned}
H_{rj}=\sqrt{\frac{1}{2}}(t+t^{'}J)
\end{aligned}
\end{equation}
Where $J^2=-1$, $t$ and $t^{'}$ are random numbers following the standard normal distribution. Based on the above, the SINR $\gamma_{rj}$ can be expressed as follows \cite{cho2010mimo}\cite{wang2017meta},
\begin{equation}
\begin{aligned}
\gamma_{rj}=\frac{P^{tran}_{rj}H^2_{rj}d^{-a}_{rj}}{N_j+\sum_{i\in \mathbf{M}, i\neq j}P_iH^2_{ri}d^{-a}_{ri} }
\end{aligned}
\end{equation}
 The data rate is denoted as $g\in\mathbb{G}$ and the decision variable $e_{rg}$ decides whether to select the data rate $g$ for user $r$,
\begin{equation}
e^{rg}=\left\{
\begin{aligned}
1,&\;\text{if data rate}\;g\;\text{is selected}\;\text{for user}\;r, \\
0,&\;\text{otherwise}.
\end{aligned}
\right.
\end{equation}
Noticing that the chosen data rate can also be written as $B_j \log_2 (1+\gamma_{rj})$, after choosing a data rate as $ge_{rg}$ for the user, the current transmit power $P^{tran}_{rj}$ can be written as follows,
\begin{equation}
\begin{aligned}
P^{tran}_{rj}=\frac{N_j+\sum_{i\in \mathbf{M}, i\neq j}P_iH^2_{ri}d^{-a}_{ri}}{H^2_{rj}d^{-a}_{rj}}(2^{\frac{ge_{rg}}{B_j}}-1)
\end{aligned}
\label{transpower}
\end{equation}
Note that $2^{\frac{ge_{rg}}{B_j}}=(1-e_{rg})+e_{rg}2^{\frac{g}{B_j}}$ and should satisfy the following constraint to ensure that a user could only select one data rate,
\begin{equation}
\begin{aligned}
\sum_{g\in\mathbb{G}}e_{rg}=1, \forall r\in\mathbb{R}
\end{aligned}
\label{cons_h7}
\end{equation}

\subsection{Latency, Power Consumption and Quality of Perception Experience}
Similar to our previous work in \cite{huang2021proactive}, the MAR service can be decomposed into computational intensive and storage intensive functionalities and are defined as $\eta$ and $\varrho$ respectively. For these functionalities, their corresponding execution locations are then denoted as $x_{ri}$ and $y_{ri}$ respectively \cite{huang2021proactive}. In a mobility event, the user's moving probability from the starting location to an allowable destination could be known to mobile operators through learning from the historical data and hence is defined as $u_{f(r)k} \in [0,1]$ ($\{f(r),k\} \subset \mathbb{M}$). The size of foreground interactions is denoted as $F^{fore}_{\eta r}$, the size of pointers used for matching AROs is denoted as $F_{\varrho r}$ and the size of the related model $s$ used for background content checking is $F^{back}_{sr}$ \cite{huang2021proactive}\cite{guo2020adaptive}. During the matching and background content checking process, the target AROs or background content are possibly not pre-cached in the local cache and such case is known as a "cache miss" (otherwise there is a "cache hit"). A cache miss in the local cache inevitably triggers the redirection of the request to the metaverse region stored in a core cloud deeper in the network and this extra cost in latency is defined as the penalty $D$. After rendering, the model and target AROs are integrated into a compressed final frame for transmission and its compressed size is denoted as $F^{res}_{sr}$.

In this section, a joint optimization scheme is proposed that aims to balance  the service delay and the energy consumption under the constraint of the user perception quality of the decomposed Metaverse AR services in the EC supported network. The cache hit/miss is expressed by the decision variable $z_{rj}$ and can be written as follows,
\begin{equation}
z_{rj}=\left\{
\begin{aligned}
1 ,&\;\text{if} \sum_{l \in \mathbf{L_{rs}}}\sum_{s \in \mathbf{S_r}}p_{sj}h^s_{rl} \geqslant L_{rs}, \\
0 ,&\;\text{otherwise}.
\end{aligned}
\right.
\end{equation}

The cache capacity of an EC and the cache hit/miss relation can be written as follows,
\begin{equation}
\begin{aligned}
&\sum_{r \in \mathbf{R}}\sum_{l \in \mathbf{L_{rs}}}\sum_{s \in \mathbf{S_r}} p_{sj}h^s_{rl}O_{l}\leq \Theta_j,\forall j\in \mathbf{M},\; j\neq j_r\\
\end{aligned}
\label{cons_h5}
\end{equation}
\begin{equation}
\begin{aligned}
&\sum_{l  \in \mathbf{N}}\sum_{s \in \mathbf{S_r}} h^s_{rl} + \epsilon \leq L_{rs} + U(1-q_{rj}) \\ &\forall j\in \mathbf{M},\; j\neq j_r,  r \in \mathbf{R}\\
\end{aligned}
\label{cons_h6}
\end{equation}
in the above expressions $\Theta_j$ denotes the cache available memory at node $j$. In \eqref{cons_h6}, to transfer the either-or constraint (i.e., $\sum_{l  \in \mathbf{N}}\sum_{s \in \mathbf{S_r}} h^s_{rl} < L_{rs} $ or $z_{rj}=1$) into inequality equations, we bring in $\epsilon$ as a small tolerance value, $U$ as a large arbitrary number and $q_{rj}$ as a new decision variable satisfying $1-q_{rj}=z_{rj}$ \cite{huang2021proactive}. Undoubtedly, increased levels of pro-caching decisions related to the background models and embedded AROs in a request inevitably brings about an extra execution burden for the matching function. Taking the above into account, the actual processing delay of the computational intensive function can be expressed as follows,
\begin{equation}
\begin{aligned}
V_{rj}=\frac{\omega_{\eta} F^{fore}_{\eta r}}{f_V^j}
\end{aligned}
\end{equation}
Similarly, the processing delay of the matching and background content checking function is assumed happen only at serves and can be written as,
\begin{equation}
\begin{aligned}
W_{rj}=\frac{\omega_{\varrho} (F_{\varrho r}+\sum_{l \in \mathbf{L_{rs}}}\sum_{s \in \mathbf{S_r}}p_{sj}h^s_{rl}O_l+\sum_{s \in \mathbf{S_r}}F^{back}_{sr}p_{sj})}{f_V^j}
\end{aligned}
\end{equation}
where $\omega_{\eta}$ and $\omega_{\varrho}$ (cycles/bit) represent the computation load of foreground interaction and background matching, $f^j_V$ is the virtual CPU frequency (cycles/sec), $F_{\varrho r}$ are the size of uploaded pointers of AROs in foreground interactions \cite{huang2021proactive}\cite{guo2020adaptive}. 
When finding the target AROs during matching, their pointers included by foreground interactions should also be transferred to the metaverse for updating. Finally, the final frame integrating the model and target AROs of is transmitted back to the user. Hence, the wired transmission delay for each user after processed by functions can be written as follows,
\begin{equation}
\begin{aligned}
\sum_{s \in \mathbf{S_r}}\sum_{j \in \mathbf{M},j\neq j_r}(C_{jA(f(r))}+C_{jA(k)})p_{sj}+\\
(C_{A(f(r))f(r)}+ \sum_{k \in \mathbf{K}}C_{A(k)k}u_{f(r)k})
\end{aligned}
\end{equation}

Note that the product of decision variables, $p_{sj}h^s_{rl}$, $p_{sj}y_{rj}$ and $p_{sj}h^s_{rl}y_{rj}$, create a non-linearity. $p_{sj}h^s_{rl}$ and $p_{sj}y_{rj}$ appear directly while $p_{sj}h^s_{rl}y_{rj}$ appears in $W_{rj}y_{rj}$, which represents the execution of the matching function at the location $j$ ($j\in \mathbb{M},j\neq j_r$).  To express the optimization problem in a nominal linear programming setting, we linearize the above expressions via new auxiliary decision variables. To this end, a decision variable $\alpha_{rsj}$ is introduced as $\alpha_{rsj}=p_{sj}y_{rj}$ and the constraints should be added as follows,
\begin{equation}
\begin{aligned}
 &\alpha_{rsj} \leqslant p_{sj}, \\ 
 &\alpha_{rsj} \leqslant y_{rj},\\
 &\alpha_{rsj} \geqslant p_{sj}+y_{rj}-1 \label{a1}\\
\end{aligned} 
\end{equation}

Similarly, a new decision variable $\beta_{rslj}$ is introduced as $\beta_{rslj}=p_{sj}h^s_{rl}$ and the constraints should be added as follows, 
\begin{equation}
\begin{aligned}
 &\beta_{rslj} \leqslant p_{sj},\\ 
 &\beta_{rslj} \leqslant h^s_{rl} ,\\ 
 &\beta_{rslj} \geqslant p_{sj}+h^s_{rl}-1 \label{a2}\\
\end{aligned}
\end{equation}

The aforementioned constraint \eqref{cons_h4} is affected and should be rewritten as follows,
\begin{subequations}
\begin{align}
h^s_{rl} \leqslant \sum_{j \in \mathbf{M}} \beta_{rslj},\;\forall r\in \mathbf{R},\;\forall s\in \mathbf{S_r},\; \forall l\in  \mathbf{L_{rs}}   \tag{\ref{cons_h4}$'$} \label{cons_h4_new} 
\end{align}
\end{subequations}

Also, note that $p_{sj}$ is a binary decision variable and causes $p_{sj}=p^2_{sj}$. Thus, we have $p_{sj}h^s_{rl}y_{rj}=\alpha_{rsj}\beta_{rslj}$. A new decision variable $\lambda_{rslj}$ is introduced as $\lambda_{rslj}=\alpha_{rsj}\beta_{rslj}$ and the following constraints should be added as follows, 
\begin{equation}
\begin{aligned}
&\lambda_{rslj} \leqslant \alpha_{rsj},\\& \lambda_{rslj} \leqslant \beta_{rslj},\\&\lambda_{rslj} \geqslant \alpha_{rsj}+\beta_{rslj}-1 \label{a3}\\
\end{aligned}
\end{equation}

Hence, the product $W_{rj}y_{rj}$ can be rewritten as follows,
\begin{equation}
\begin{aligned}
\frac{\omega_{\varrho} (F_{\varrho r}y_{rj}+\sum_{l \in \mathbf{L_{rs}}}\sum_{s \in \mathbf{S_r}}\lambda_{rslj}O_l+\sum_{s \in \mathbf{S_r}}F^{back}_{sr}\alpha_{rsj})}{f_V^j}
\label{storage_intensive_comp}
\end{aligned}
\end{equation}

By checking whether users share the same metaverse region by $A(f(t))=A(f(r)), \{t,r\}\subset\mathbb{R}$, we can ensure the user could also view other updates happening in the same metaverse region. Based on the previous modelling of wireless channel, the wireless transmission delay in a mobility event can be written as follows,
\begin{equation}
\begin{aligned}
&\sum_{r\in \mathbf{R}}\frac{F^{fore}_{\eta r}+\sum_{t\in\mathbf{R}, A(f(t))=A(f(r))}\sum_{s\in\mathbf{S_r}}p_{sj}F^{res}_{st}}{ge_{rg}}+\\&\sum_{r\in \mathbf{R}}\sum_{k\in\mathbf{K}}u_{f(r)k}\frac{F^{fore}_{\eta r}+\sum_{t\in\mathbf{R}, A(t)=A(k)}\sum_{s\in\mathbf{S_r}}p_{sj}F^{res}_{st}}{ge_{rg}}
\label{wireless}
\end{aligned}
\end{equation}

Noticing that with constraint \eqref{cons_h7}, $\frac{1}{e_{rg}}$ can be replaced by $e_{rg}$ for linearization. By introducing a new decision variable $\phi_{rlsg}$ with following constraints,
\begin{equation}
\begin{aligned}
&\phi_{rsg} \leqslant e_{rg},\\&\phi_{rsg} \leqslant p_{sj},\\&\phi_{rsg} \geqslant e_{rg}+p_{sj}-1 \label{a4}\\
\end{aligned}
\end{equation}

Thus, the previous formula \eqref{wireless} can be updated as follows,
\begin{equation}
\begin{aligned}
&\frac{1}{g}\sum_{r\in \mathbf{R}}(1+\sum_{k\in\mathbf{K}}u_{f(r)k}) (F^{fore}_{\eta r}e_{rg}+\\&\sum_{t\in\mathbf{R}, A(f(t))=A(f(r))}\sum_{s\in\mathbf{S_r}}\phi_{rsg}F^{res}_{st} )
\label{wireless_new}
\end{aligned}
\end{equation}

Based on the above derivations and inline with \cite{huang2021proactive}, the overall latency can be written as follows,
\begin{equation}
\begin{aligned}
L=&\eqref{wireless_new}+\sum_{r \in \mathbf{R}}\sum_{i \in \mathbf{M}}(C_{f(r) i}+V_{ri})x_{ri}+\\
&\sum_{r \in \mathbf{R}}\sum_{i \in \mathbf{M}}\sum_{j \in \mathbf{M},j\neq j_r} (\eqref{storage_intensive_comp}+ C_{ij}\xi_{rij}+C_{A(f(r))f(r)}+\psi_{rj}D  )+\\ \\
&\sum_{s \in \mathbf{S_r}}\sum_{j \in \mathbf{M},j\neq j_r}(C_{jA(f(r))}+C_{jA(k)})p_{sj}+ \\
&\sum_{r \in \mathbf{R}}\sum_{k \in \mathbf{K}}(C_{A(k)k}+C_{ki}x_{ri})u_{f(r)k}
\end{aligned}
\end{equation}
where $V_{ri}$ is the processing delay of computational intensive function \cite{huang2021proactive}. $L_{max}$ here denotes the maximum allowed service latency and has $\frac{L}{L_{max}} \in [0,1]$.

The energy efficiency of the system during each service time slot is measured by the production of its total power and running time. The server total power consists of the transmission power and CPU processing power at target ECs. Denote the required CPU processing power of the user $r$ at the node $j$ as $P^{cpu}_{rj}$ and the CPU chip architecture coefficient as $k_0$ (e.g. $10^{-18}$) \cite{dong2019deep}. Then the power at the EC can be achieved through $k_0(f^j_V)^2$ (J/s) based on measurements in \cite{zhang2013energy}\cite{miettinen2010energy}. Noticing that for both conditions, the background contents are processed at servers, hence the consumed processing time of the server is,
\begin{equation}
\begin{aligned}
T_{cpu} =\sum_{r\in\mathbf{R}}\sum_{j\in\mathbf{M},j\neq m_r}V_{rj}x_{rj}+\sum_{r\in\mathbf{R}}\sum_{j\in\mathbf{M}}W_{rj}y_{rj}
\end{aligned}
\end{equation}
The wireless transmission happens no matter if executing foreground interactions at terminals. Since the selected data rate is $ge_{rg}$, the wireless transmission time is,
\begin{equation}
\begin{aligned}
T_{tran} =\sum_{r\in\mathbf{R}}\sum_{j\in\mathbf{M},j\neq j_r}\frac{F^{fore}_{\eta r}}{ge_{rg}}  + \sum_{r\in\mathbf{R}}\sum_{j\in\mathbf{M},j= j_r}\frac{F_{\varrho r}}{ge_{rg}} 
\end{aligned}
\end{equation}
Finally, through the production of power and time, the total consumed energy at the server side can be written as follows,
\begin{equation}
\begin{aligned}
E_{server} &=\sum_{r\in\mathbf{R}}\sum_{j\in\mathbf{M}}(P^{tran}_{rj}T_{tran}+P^{cpu}_{rj}T_{cpu})\\
&=\sum_{r\in\mathbf{R}}\sum_{j\in\mathbf{M}}(\frac{N_j+\sum_{i\in \mathbf{M}, i\neq j}P_iH^2_{ri}d^{-a}_{ri}}{H^2_{rj}d^{-a}_{rj}}(2^{\frac{ge_{rg}}{B_j}}-1))\\
&\quad (\sum_{j\neq j_r}\frac{F^{fore}_{\eta r}}{ge_{rg}}  + \sum_{j= m_r}\frac{F_{\varrho r}}{ge_{rg}})\\
&\quad +\sum_{r\in\mathbf{R}}\sum_{j\in\mathbf{M}}k_0(f^j_V)^2 (W_{rj}y_{rj}+\sum_{j\neq j_r}V_{rj}x_{rj}) 
\end{aligned}
\end{equation}

At the terminal, this paper also follows the Amdahl’s law to model its energy consumption, in which considers the parallelism by the serial and parallel portion respectively \cite{amdahl1967validity}. In this paper, the metaverse AR functions work by serial and the parallel portion is assumed to be 0 for simplicity like \cite{song2019energy}. As mentioned earlier, the dynamic foreground interactions including some highly used AROs could be proactively cached at terminals with the matching function. The 3D background model, on the other hand, is much larger and might serve multiple users in a region. Hence, it is not recommended to be stored or processed at the terminal. In addition, the metaverse application should not exceed a certain portion of the whole terminal cpu resources so that other functionalities could work properly \cite{song2019energy}. Denoting the consumed portion as $\Gamma_r \in [30\%, 50\%]$ \cite{song2019energy}, then $\frac{V_{rj}x_{rj}}{\Gamma_r}$ means processing metaverse AR functions at the terminal requires a longer time. The Energy consumption of terminals could be written as follows,
\begin{equation}
\begin{aligned}
E_{terminal} =&P_{terminal}T_{terminal}\\
=&\sum_{r\in\mathbf{R}}\sum_{j=j_r}k_0(f^j_V)^2 \frac{V_{rj}x_{rj}}{\Gamma_r}
\end{aligned}
\end{equation}

Finally, the overall system energy consumption is $E=E_{server}+E_{terminal}$.
$E_{max}$ represents the maximum possible energy consumption of the system. It also has $\frac{E}{E_{max}} \in [0,1]$.

SSIM is applied to reveal the quality of perception experience. In this paper, the video coding scheme (e.g. H.264) and frame resolution (e.g. 1280$\times$720) are assumed as pre-defined \cite{kato2021split}. Then SSIM is mainly affected by data rate and a concave function could be applied to reveal the relation between them\cite{kato2021split}. Hence, denote the set of SSIM values for each ARO under corresponding data rates as $\mathbb{SSIM}_{l}, l\in\mathbb{L}_r$. The overall quality of perception experience $Q$ can be written as follows,
\begin{equation}
\begin{aligned}
Q=\sum_{r\in\mathbf{R}}\sum_{l\in\mathbf{L_r}}\sum_{g\in\mathbf{G}}\sum_{c\in\mathbf{SSIM}_{l}} e_{rg}c 
\end{aligned}
\end{equation}
To maintain the user experience above an acceptable level, the perception quality constraint could be added as follows,
\begin{equation}
\begin{aligned}
\frac{Q}{Q_{max}}\geq Q_{bound}
\label{cons_quality}
\end{aligned}
\end{equation}
where $Q_{max}$ is the maximum quality through selecting max allowable data rate and storing as many AROs as possible.

We denote the weight parameter is denoted as $\mu \in [0,1]$ and the joint optimization problem can eventually be written as follows,
\begin{subequations}
\begin{align}
\mathop{min}
&\; \mu \frac{L}{L_{max}}+(1-\mu) \frac{E}{E_{max}}
\label{JOP:1}
\\
\nonumber
\\
\text{s.t.}\;& z_{rj} = 1-q_{rj}, \: \forall j\in \mathbf{M},  r \in \mathbf{R}
\label{JOPcon:1}
\\
&\sum_{r \in \mathbf{R}} (x_{rj}+y_{rj})\leq \Delta_j,\forall j\in \mathbf{M},j\neq j_r
\label{JOPcon:2}
\\
&\sum_{j  \in \mathbf{M}} x_{rj}=1,\forall r\in \mathbf{R}
\label{JOPcon:3}
\\
&\sum_{j  \in \mathbf{M},j\neq j_r} y_{rj} =1,\forall r \in \mathbf{R}
\label{JOPcon:4}
\\
&\xi_{rij} \leq x_{ri}, \: \forall r \in \mathbf{R}, i,j \in  \mathbf{M},j\neq j_r
\label{JOPcon:5}
\\
&\xi_{rij} \leq y_{rj}, \: \forall r \in \mathbf{R}, i,j \in  \mathbf{M},j\neq j_r
\\
&\xi_{rij} \geq x_{ri} + y_{rj} -1, \: \forall r \in \mathbf{R}, i,j \in  \mathbf{M},j\neq j_r
\\
&\psi_{rj} \leq z_{rj}, \: \forall r \in \mathbf{R}, j \in  \mathbf{M},j\neq j_r
\\
&\psi_{rj} \leq y_{rj}, \: \forall r \in \mathbf{R}, j \in  \mathbf{M},j\neq j_r
\\
&\psi_{rj} \geq z_{rj} + y_{rj} -1, \: \forall r \in \mathbf{R}, j \in  \mathbf{M},j\neq j_r
\label{JOPcon:10}
\\
&x_{rj}, y_{rj},p_{sj}, h^s_{rl}, z_{rj}, q_j\in  \{0,1\},\nonumber\\
&\alpha_{rsj},\beta_{rslj},\lambda_{rslj}, \phi_{rslg},\psi_{rj}, \xi_{rij}\in \{0,1\}, \nonumber\\
&\forall r \in \mathbf{R},  j\in \mathbf{M},  l \in \mathbf{L_{rs}}, s\in \mathbf{S_r}
\label{JOPcon:11}
\\
&\eqref{cons_h1},\;\eqref{cons_h2},\;\eqref{cons_h3},\;\eqref{cons_h4_new},\;\eqref{cons_h7},\;\eqref{cons_h5},\;\eqref{cons_h6}, \nonumber \\
&\eqref{a1},\;\eqref{a2},\;\eqref{a3},\;\eqref{a4},\;\eqref{cons_quality}\nonumber
\end{align}
\end{subequations}

As mentioned earlier, any assignment relating to the storage intensive functions ($y_{rj}$) are limited to only ECs and hence should apply the constraint $j\neq j_r$. The constraint \eqref{JOPcon:1} together with constraints \eqref{cons_h1} to \eqref{cons_h4_new} reveal the relation between pre-caching decisions and the cache miss/hit for each request \cite{huang2021proactive}. The constraint \eqref{JOPcon:2} is the virtual machine limitation while \eqref{JOPcon:3} and\eqref{JOPcon:4} guarantee the once execution of each function of a request at a single server as explained in \cite{huang2021proactive}. The constraints \eqref{a1} to \eqref{a3} and \eqref{JOPcon:5} to \eqref{JOPcon:10} are auxiliary and required to solve the product of decision variables for linearization.

\section{Numerical Investigations}
In this section, the effectiveness of the proposed optimization scheme, which will be referred to as Optim in the sequel, is investigated %via 50 monte carlo simulations 
and compared with a number of nominal/baseline schemes.
\begin{figure}[htbp]
\centering
\includegraphics[width=0.9\linewidth,height=0.5\linewidth]{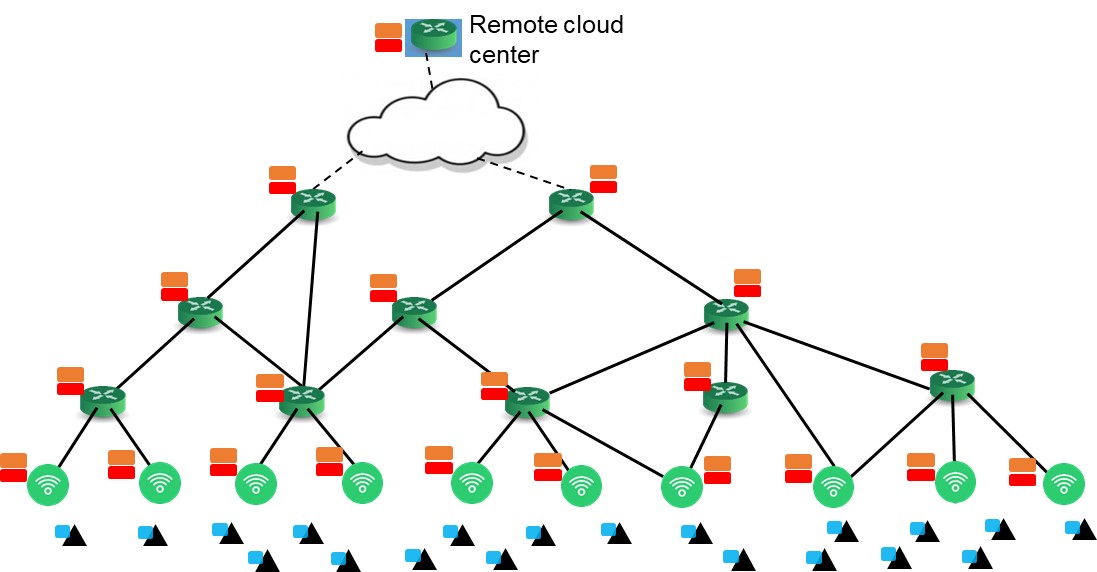} 
\caption{A typical tree-like designed network topology}
\label{fig:topology}
\end{figure}
%Same as in \cite{huang2021proactive}, 
A nominal tree-like network topology shown as \eqref{fig:topology} is applied with 20 ECs in total and 6 ECs being activated for the current metaverse AR service and 30 requests are sent by MAR devices. The remaining available resources allocated for metaverse AR support within an EC are assumed to be CPUs with frequency as 4 to 8 GHz, CPU chip architecture coefficient as $10^{-18}$ (affected by the chip's design and structure) \cite{dong2019deep}, 4 to 8 cores and $[100,400]$MBytes of cache memory \cite{huang2021proactive}. Similarly, the mobiles AR devices are assumed to own a CPU with around 1 GHz frequency, 4 cores and $[0,100]MBytes$ available cache memory for Metaverse AR applications \cite{song2019energy}. According to \cite{chen2018understanding}, the power of AR applications should not exceed 50\% of the mobile device's CPU total power of the mobile device ($2-3W$) so that other functionalities could operate efficiently. Hereafter, a nominal frame rate is asssumed as 15 frames/second and the rendering happens at every other frame (\~ 133.2ms interval) \cite{cozzolino2022nimbus}\cite{niu2018learning}\cite{naman2013inter}. Thus, the service delay of the aforementioned work flow within the interval could be regarded as acceptable. Each request requires a single free resource unit for each service function, such as for example a Virtual Machine (VM) \cite{liu2018edge}. Up to 14 available VMs are assumed in each EC, with equal splitting of the available CPU resources \cite{huang2021proactive}. Note that different view ports lead to different models of the metaverse \cite{guo2020adaptive} and up to 4 different models can be cached. All target AROs must be integrated with the corresponding model and rendered within the frame before being streamed to the end user based on a matched result. After triggering the metaverse MAR service, pointers to identify AROs like a name or index are usually a few bytes  \cite{zhang2021multi} and hence their transmission and processing are neglected in the following simulations. The set of available data rates is $\{2,3,...,8\}$Mbps and its corresponding SSIM values set is $\{0.955, 0.968,...,0.991\}$ \cite{kato2021split}. We require the acceptable average SSIM above 0.97 ($Q_{bound}$). For a nominal 5G base station, we assume its cell radius as 250m, its carrier frequency as 2GHz, its transmit power as 20dBm, the noise power as $10^{-11}$W, the path loss exponent is 4, its maximum resource blocks as 100 and without the loss of generality, each user can utilize only one resource block \cite{korrai2019slicing}\cite{li2020computing}\cite{chettri2019comprehensive}. As mentioned earlier, we accept a predefined video coding scheme H.264 with a fixed frame resolution as 1280$\times$720 \cite{kato2021split} in RGB (8 bits per pixel). Based on the given resolution, the size of foreground interactions after decoding and compressing can be calculated through multiplying the coefficients $\frac{5}{9}$ and $10^{-3}$ \cite{guo2020adaptive}. Matlab on a personal PC with its CPU of intel i7, 6500U and 2 cores is applied for the simulation. Key simulation parameters are shown below in Table \eqref{tab:parameters}.
\begin{table}[ht]
\caption{Simulation parameters}
\begin{center}
\begin{tabular}{c|c}
\hline
\textbf{Parameter}&\textbf{Value} \\
\hline
Number of available ECs&$6$\\
\hline
Number of available VMs per EC (EC Capacity)&$14$\\
\hline
Number of requests&$30$\\
\hline
Number of available models per user &$4$\\
\hline
AR object size&$(0,10]$ MByte\\
\hline
Total moving probability&$[0,1]$\\
\hline
Cell Radius&$250$m\\
\hline
Remained Cache Capacity per EC&$[100,400]$ MByte\\
\hline
EC CPU frequency&[4,8] GHz\\
\hline
EC CPU cores&[4,8]\\
\hline
EC CPU core portion per VM&$0.25-0.5$\\
\hline
Remained Cache Capacity per Terminal&$[0,100]$ MByte\\
\hline
Terminal CPU frequency&$1$ GHz\\
\hline
Terminal CPU cores&$4$\\
\hline
CPU Architecture Coefficient&$10^{-18}$ \\
\hline
Foreground Interaction Computational Load&$4$ cycles/bit\\
\hline
Background Content Checking Computational Load &$10$ cycles/bit\\
\hline
Carrier Frequency &$2$ GHz\\
\hline
Transmission Power &$20$dBm\\
\hline
Path Loss Exponent&$4$\\
\hline
Noise Power&$10^{-11}$W\\
\hline
Number of Resource Blocks&$100$\\ 
\hline
Frame Resolution&1280$\times$720\\
\hline
Average latency per hop &$2$ ms\\
\hline
Cache miss penalty&$25$ ms\\
\hline
\end{tabular}
\label{tab:parameters}
\end{center}
\end{table}

In following figures and discussion, the optimized scheme considering terminals is denoted as OptimT while the other one without terminals is denoted as as OptimNT. The OptimNT scheme could be regarded as an natural extension from our previous work \cite{huang2022mobility}. Two other baseline schemes sharing same caching decisions as the proposed Optim scheme are also implemented for comparison. Those are the Random Selection Scheme (RandS) and the Closest EC First Scheme (CEC) \cite{tocze2019orch}. The RandS scheme operates a random EC selection while the other two both select the closest EC to the user's initial location. The CEC scheme also accepts the second closest one as a back up choice \cite{tocze2019orch}. 

\begin{figure}[ht]
\centering
\includegraphics[width=0.9\linewidth]{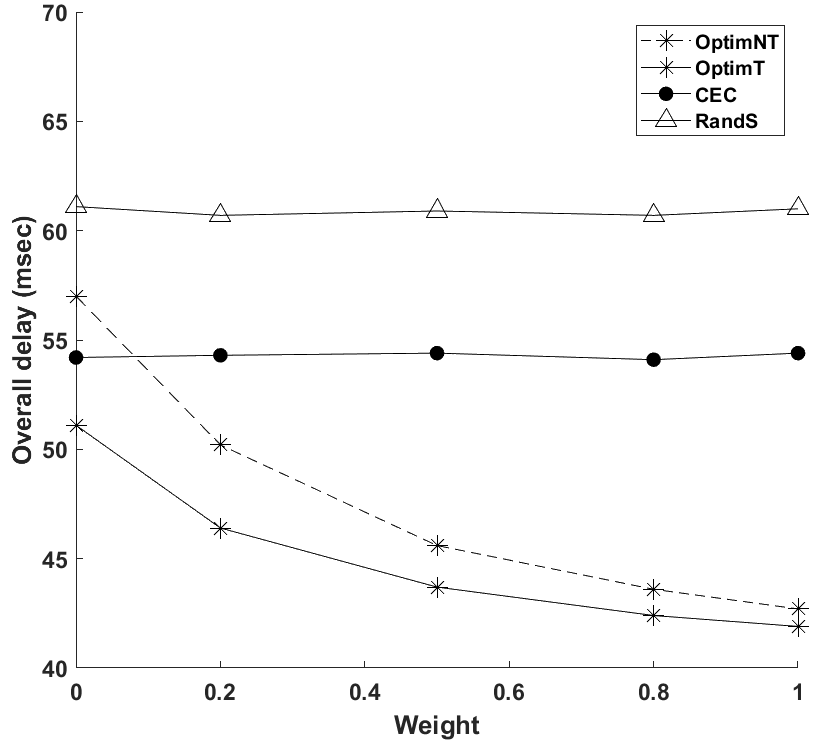}
\caption{Overall Delay with Weight $\mu$ (6 EC, 30 Requests, EC Capacity is 14 and total mobility probability is 1)}
\label{fig:W&D}
\end{figure}
\begin{figure}[ht]
\centering
\includegraphics[width=0.9\linewidth]{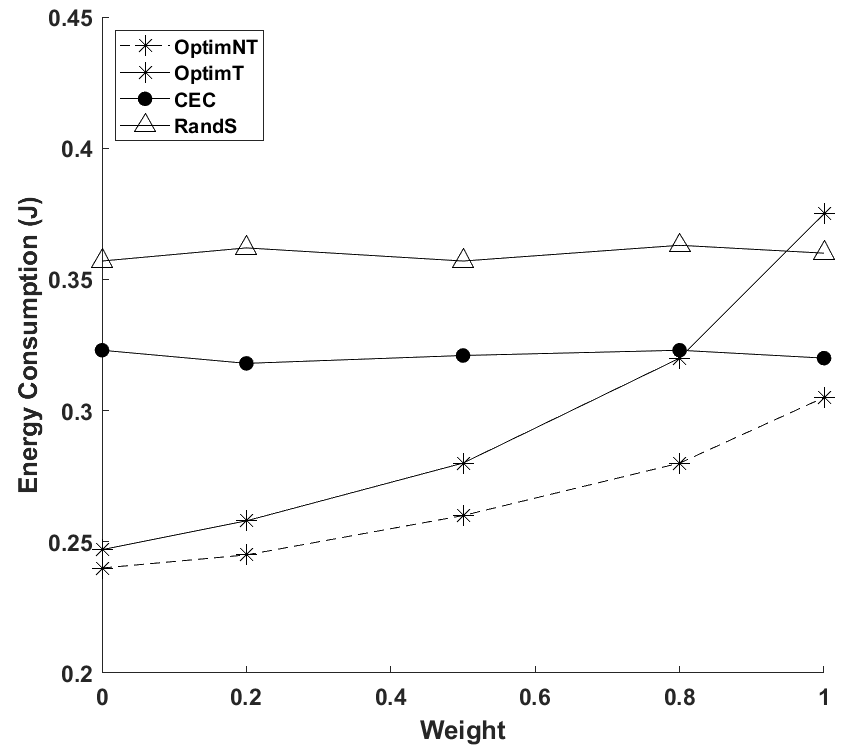}
\caption{Average Energy Consumption with Weight $\mu$}
\label{fig:W&E}
\end{figure}
According to Fig. \eqref{fig:W&D}, the service delay for each request of proposed schemes drop as expected with an increasing weight $\mu$. With a larger weight, the proposed schemes tend to select a larger data rate and direct the service to more powerful ECs, which naturally lead to a smaller overall delay. Compared to the OptimNT scheme, for example, the gain in delay of the Optim scheme ranges from 1.9\% to 10.4\%. When seeking for the best energy efficiency ($\mu=0$), since the CPU resources at the terminals are also shared by other functionalities, the OptimT scheme also tries to avoid the occupation of terminals. Hence, in this case, these two schemes share similar solutions and approach to each other. Noticing that the proposed schemes do not care about latency cost, they could choose further and busy EC which even cause the OptimNT scheme become worse than the CEC scheme. Afterwards, as the weight $\mu$ increases and the emphasis is placed on latency rather than energy,  causes the OptimT scheme to allocate some foreground interactions at terminals and becomes better than OptimNT. Since the baseline schemes neglect energy consumption, service decomposition and mobility, their gaps to the proposed schemes become larger with an increasing weight. However, such gain in delay comes with an extra cost in energy consumption. As shown by Fig. \eqref{fig:W&E}, the energy consumption per request of the proposed schemes increase with a larger weight. Compared to the OptimNT scheme, the OptimT scheme consumes 2.9\% to 23.0\% more energy under different weights. Thus, it might not always be worthy to endure lots of energy consumption for narrow gains in delay. By selecting a suitable weight, a balance could be achieved with the OptimT scheme between delay and energy consumption. Through the utilization of terminals, the OptimT scheme becomes the most sensitive to the energy consumption and there could be instances that it might consume more energy that the RandS scheme.

\begin{figure}[ht]
\centering
\includegraphics[width=0.9\linewidth]{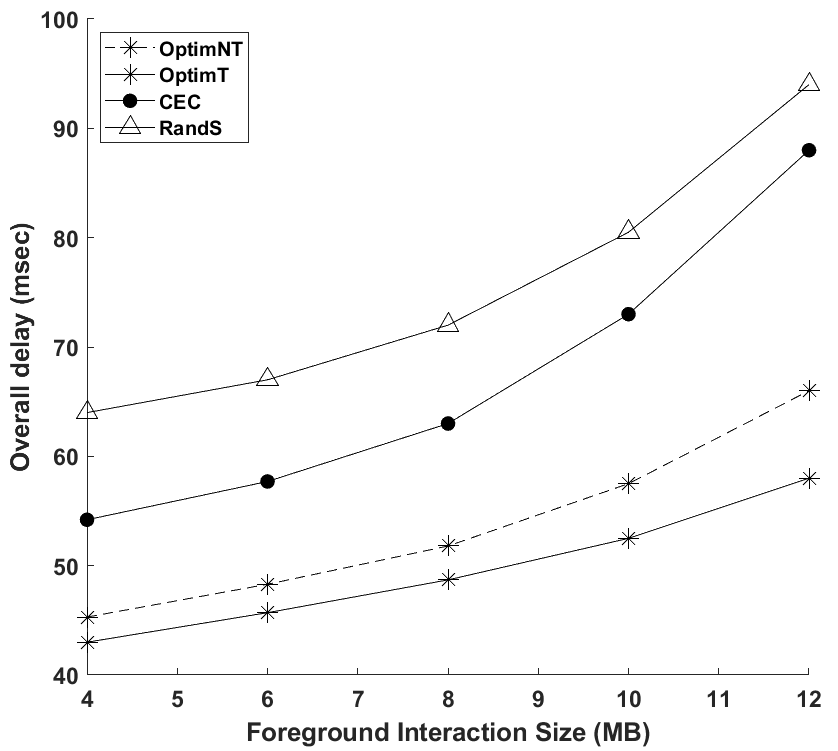}
\caption{Overall Delay with Foreground Interaction Size (6 EC, 30 Requests, $\mu=0.5$, EC Capacity is 14 and total mobility probability is 1)}
\label{fig:Fore&D}
\end{figure}
\begin{figure}[ht]
\centering
\includegraphics[width=0.9\linewidth]{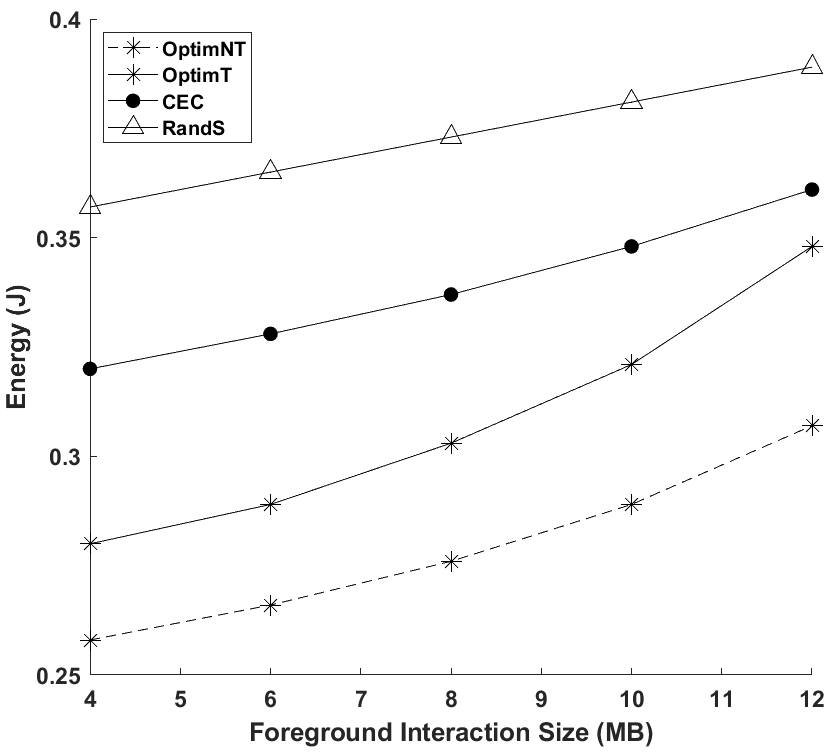}
\caption{Average Energy Consumption with Foreground Interaction Size (6 EC, 30 Requests, $\mu=0.5$, EC Capacity is 14 and total mobility probability is 1)}
\label{fig:Fore&E}
\end{figure}
\begin{figure}[ht]
\centering
\includegraphics[width=0.9\linewidth]{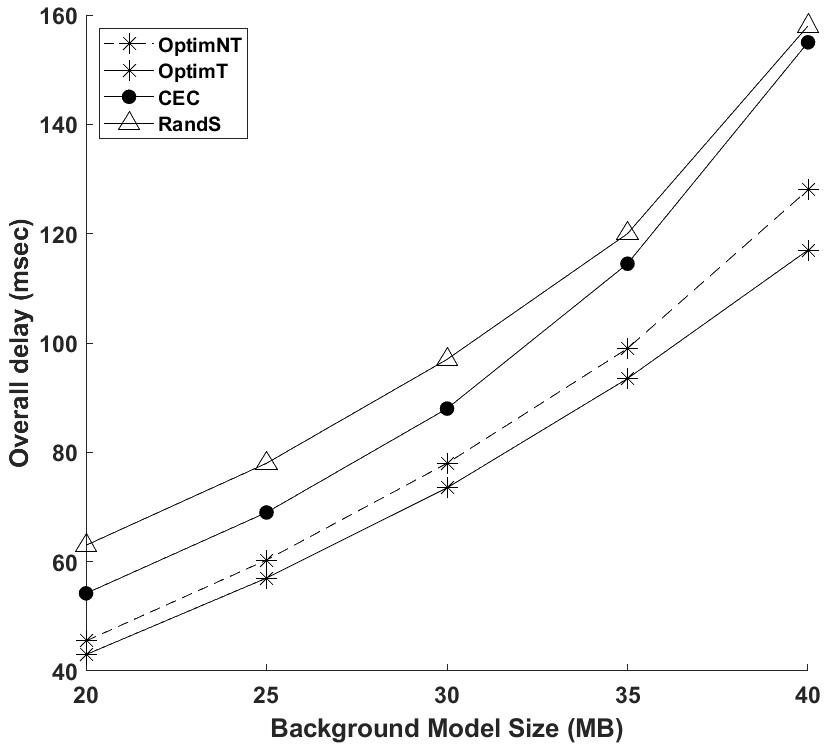}
\caption{Overall Delay with Background Model Size (6 EC, 30 Requests, $\mu=0.5$, EC Capacity is 14 and total mobility probability is 1)}
\label{fig:Back&D}
\end{figure}
Clearly, the user experience could be elevated through viewing more AROs in foreground interactions or more delicate scenes from background models. Fig. \eqref{fig:Fore&D} reveals the variation of delay with the increasing foreground interaction size.When the average foreground interaction size is not too large and there are still enough resources at target ECs, OptimNT and baseline schemes increase almost linearly at a similar speed. When the size keeps increasing and resources become limited, the CEC scheme becomes the most sensitive one because it only targets on several closest ECs and is easier to trigger the penalty. The optimT scheme, on the other hand, maintains the least latency and the least increasing tendency. To this end, it could save up to 13.8\% and 51.7\% delay compared respectively to the OptimNT scheme and the CEC scheme. Fig. \eqref{fig:Fore&E} further reveals the variation of energy in this case. Baseline schemes process foreground interactions at ECs without considering energy. Hence, their decisions are not obviously affected by the size of foreground interactions and their energy consumption increases almost linearly. The OptimNT schemes keeps finding more suitable ECs according to current foreground interaction size and remained resources while the OptimT schemes further enables terminals to process some works. Compared to the CEC scheme, they could save over 14.3\% energy. It is necessary to point out that the energy consumption will not be taken into account when redirecting the request to the further core cloud and triggering the penalty. According to Fig. \eqref{fig:Back&D}, the background model size is much larger and could cause a significant increase in delay. Note that the terminals could take charge of some foreground interactions and hence could make room for background models in the OptimT scheme. It is still the best scheme in terms of delay and could be up to 9.8\% less than the OptimNT scheme. As mentioned earlier, the proactively caching and processing of background models only happen at ECs, these schemes share a similar level of increased energy consumption until triggering the overloading penalty.       

\begin{figure}[ht]
\centering
\includegraphics[width=0.9\linewidth]{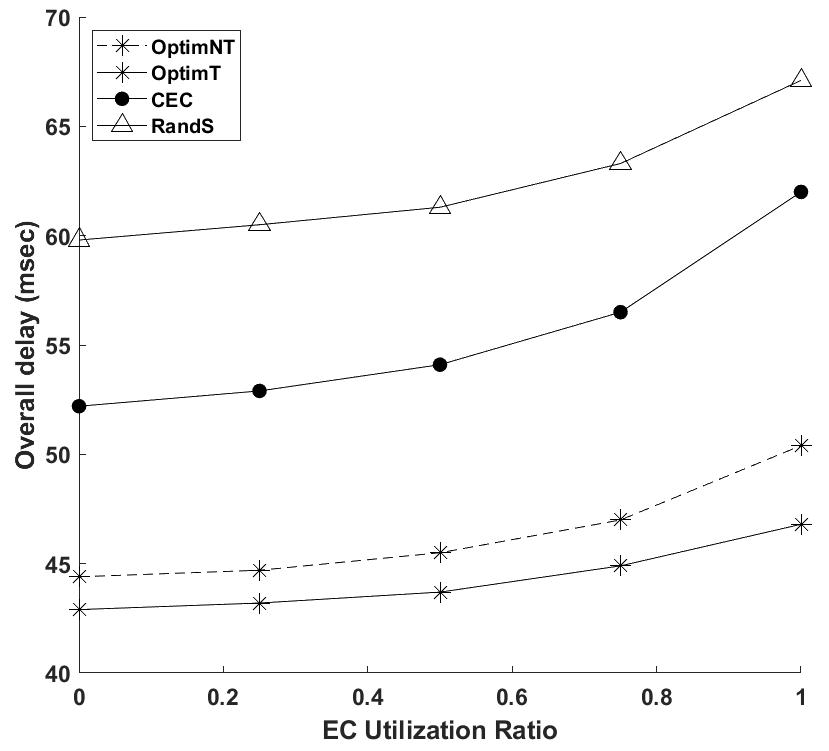}
\caption{Overall Delay with Average EC Utilization Rate (6 EC, $\mu=0.5$ and total mobility probability is 1)}
\label{fig:U&D}
\end{figure}
\begin{figure}[ht]
\centering
\includegraphics[width=0.9\linewidth]{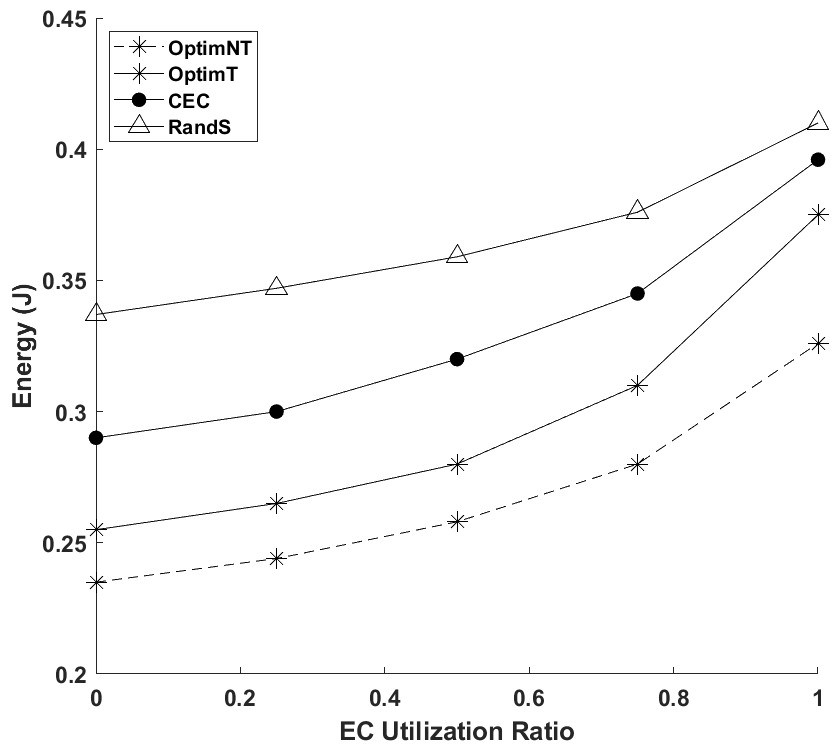}
\caption{Energy Consumption with Average EC Utilization Rate (6 EC, $\mu=0.5$ and total mobility probability is 1)}
\label{fig:U&E}
\end{figure}

The number of available VMs activated in an EC is known as the EC capacity. For a given EC capacity (e.g. 14), the the ratio between the different number of requests (e.g. [30,40]) and the EC capacity could be applied to represent the average EC Utilization in the network. Then, this rate is normalized into [0,1] for better presentation. As shown in Fig. \eqref{fig:U&D} and \eqref{fig:U&E}, the increasing EC Utilization rate indicates a more congested network and hence as expected the delay and energy consumption increase as well. Compared to the OptimT scheme, the OptimNT scheme is still more sensitive in terms of energy but better in terms of delay. Thus, its consideration of terminals benefits delay at the cost of energy. Observe from Table \eqref{tab:no_mob}, that even when there is no mobility event, the proposed OptimT scheme is still slightly better than other baseline schemes because its flexibility of terminals could better avoid potential EC overloading. Therefore, the proposed OptimT and OptimNT schemes have an obvious advantage over baseline schemes and is recommended in a congested network and a high user physical mobility scenario. Especially when the MAR terminal still owns enough energy capacity, its computing resources should not be neglected and hence the OptimT scheme is more suitable in this case.

\begin{table}[htb]
\caption{Overall Delay in no mobility event \\($\mu=1$, 6 ECs, 30 requests and EC Capacity is 14)}
    \centering
    \begin{tabular}{c|c|c|c|c|c|c}
    \hline
         \textbf{Scheme}&\textbf{OptimT}&\textbf{OptimNT}&\textbf{CFS}&\textbf{RandS}\\
         \hline
         \textbf{Delay (ms)}&38.8&40.1&40.7&60.8\\
         \hline
    \end{tabular}
    \label{tab:no_mob}
\end{table}

\section{Conclusions}
Extending MAR applications into the metaverse, it is expected to incorporate rendering and updating of high quality AR metadata in order to provide a more realistic experience. Hence, such applications are delay and energy sensitive and are demanding in caching and computing resources. In this paper, a joint optimization scheme is proposed by considering explicitly the model rendering, user mobility and service decomposition to achieve the balance between energy consumption and service delay under the constraint of user perception quality for metaverse MAR applications. Technical improvements on AR devices allow them to take over more tasks and its potential in metaverse is explored and then compared with  the terminal oblivious one in this paper. A wide set of numerical investigations reveals that the proposed terminal aware framework provides improved decision making compared to baseline schemes for energy consumption and resource allocation for metaverse MAR applications, especially under a congested network and high mobility scenario.

\bibliographystyle{IEEEtran}
\bibliography{bibliography.bib}

\end{document}